\documentstyle[11pt,fleqn]{article}
\input{psfig.tex}%\parindent=0 pt
\title{Comments on "Various Super Yang-Mills Theories with Exact
Supersymmetry on the Lattice"} % Declares the title.
\author{A. Kwang-Hua Chu \thanks{Present Address : P.O. Box 39, Tou-Di-Ban, Xihong Road,
Urumqi 830000, PR China.  }}  %address%\cite{PKU:1999}
\date{P.O. Box 30-15, Shanghai 200030, PR China}
\begin{document}           % End of preamble and beginning of text.
\maketitle
\begin{abstract}
We make corrections on the paper by Sugino [{\it JHEP} {\bf 0501}
(2005) 016].  \newline

\end{abstract}
%\pacs{}
%%\end{titlepage}      %%\twocolumn     %%\nopagebreak   %\oddsidemargin=1mm
\doublerulesep=6mm    %
\baselineskip=6mm
\bibliographystyle{plain}               %\section{Introduction}
\noindent

Sugino gave a presentation [1] in order to continue to construct
lattice super Yang-Mills theories along the line discussed in the
previous papers [2-3]. In his construction of $N$ = 2, 4 theories
in four dimensions, the problem of degenerate vacua seen in [2] is
resolved without affecting exact lattice supersymmetry, while in
the weak coupling expansion some surplus modes appear both in
bosonic and fermionic sectors reflecting the exact supersymmetry.
A slight modi- fication to the models is made such that all the
surplus modes are eliminated in two- and three-dimensional models
obtained by dimensional reduction thereof. $N$ = 4, 8 models in
three dimensions need fine-tuning of three and one parameters
respectively to obtain the desired continuum theories, while
two-dimensional models with $N$ = 4, 8 do not require any
fine-tuning. \newline Some of Sugino's results, say, in page 3 and
in page 10 for the equations (2.10), (3.5) and (3.6) are based on
the derivations in Appendix A (Resolution of Vacuum Degeneracy).
The minimum $\vec{\Phi} (x)+\Delta \vec{\Phi} (x) = 0$ is realized
by the unique configuration $U_{\mu\nu} (x) = 1$ for any choice of
the constant $r$ in the range $r=\cot(\phi)$ : $0<\phi \le
\pi/(2N)$. \newline The present author, after following his
derivations in Appendix A., would like to point out that there are
other possibilities except the unique solution $U=V=1$ which
Sugino claimed in [1]. \newline In fact, Sugino missed one other
condition : $[u_k + v_{i(k)}]/2=-\pi$! \newline To easily
illustrate our reasoning, we plot a schematic figure (shown as
Fig. 1) of which the possible  regions $\overline{A'A}$ and
$\overline{B'B}$ are related to the equation (A.9) in [1] for
$|\cos(\frac{u_k + v_{i(k)}}{2}-\phi)| \ge \cos \phi$. This
equation comes from
\begin{equation}
  \cos(\frac{u_k + v_{i(k)}}{2}-\phi) \cos(\frac{u_k - v_{i(k)}}{2}) = \cos
  \phi,  \hspace*{12mm} k=1,\cdots, N,
\end{equation}
once we impose $|\cos(\frac{u_k - v_{i(k)}}{2})|\le 1$ (follow
Sugino's reasoning in [1]), where $0<\phi \le \pi/(2N)$, and
\begin{displaymath}
 e^{i u_1} \cdots e^{i u_N}=1,  \hspace*{24mm} e^{i v_{i(1)}} \cdots e^{i
 v_{i(N)}}=1.
\end{displaymath}
There will be two solutions for N=1 : $u_1 =v_{i(1)}=\pi$; $u_2
=v_{i(2)}=-\pi$ or $u_2 =v_{i(2)}=\pi$ once $\phi=\pi/2$. Under
this case, $e^{i u_1}=e^{i\pi}=-1$ in the expression of the
equation (A.4) in [1].

\newpage

\oddsidemargin=3mm

\pagestyle{myheadings}

\topmargin=-18mm

\textwidth=17cm \textheight=26cm
%\begin{document} %\psdraft         %wn0.s      &   un0.ps
\psfig{file=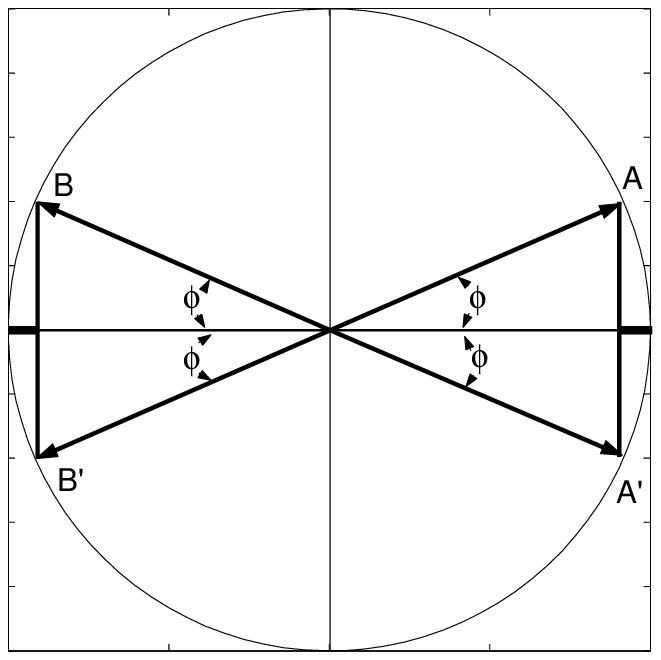,bbllx=0.1cm,bblly=10.5cm,bburx=12cm,bbury=23.8cm,rheight=9cm,rwidth=9cm,clip=}
%
%\vspace{2mm}
\begin{figure}[h]
\hspace*{3mm} Fig. 1 \hspace*{1mm} Schematic plot for the possible
region $\overline{A'A}$ and $\overline{B'B}$ for the equation
(A.9) in [1] \newline \hspace*{4mm} for $|\cos(\frac{u_k +
v_{i(k)}}{2}-\phi)| \ge \cos \phi$, $0<\phi \le \pi/(2N)$,
$k=1,\cdots, N$.
\end{figure}

\end{document}